# Nano-Focusing of Vortex Beams with Hyperbolic Metamaterials

Wenhao Li[1], Jacob LaMountain[2], Evan Simmons[2], Anthony Clabeau[3], Robel Y. Bekele[3], Jason D. Myers[3], Takashige Omatsu[4], Jesse Frantz[3], Viktor A. Podolskiy[2], Natalia M. Litchinitser[1*]

[1]Department of Electrical and Computer Engineering, Duke University; Durham, NC, 27708, USA
[2]Department of Physics and Applied Physics, University of Massachusetts Lowell; Lowell, MA, 01854, USA
[3]US Naval Research Lab; Washington, DC, 20375, USA
[4]Molecular Chirality Research Center, Chiba University; Chiba, 263-8522, Japan
* Corresponding author. Email: natalia.litchinitser@duke.edu

## Abstract

The synergy of judiciously engineered nanostructures and complex topology of light creates unprecedented opportunities for tailoring light-matter interactions on the nanoscale. Electromagnetic waves can carry multiple units of angular momentum per photon, stemming from both spin and orbital angular momentum contributions, offering a potential route for modifying the optical transition selection rules. However, the size difference between a vortex beam and quantum objects limits the interaction strength and the angular momentum exchange. Here, we demonstrate the sub-diffraction-limited focusing of a vortex beam using the high in-plane wave number modes present in hyperbolic metamaterials. The spin-orbit interaction within the hyperbolic structure gives rise to the formation of an optical skyrmion with a deep subwavelength structure, which may enable the exploration of new light-matter interaction phenomena.

## Introduction

The discovery of the OAM of light opened a new branch of optical physics that has facilitated studies of such phenomena as rotary photon drag, angular uncertainty relationships, rotational frequency shifts, and spin-orbit coupling[1-3]. The presence of the OAM can significantly change the light-matter interactions and optical transitions in emitters due to the extra angular momentum the photon carries[4]. The OAM beams (also known as vortex beams) have been proposed and employed in many applications such as high dimensional quantum information[5], optical tweezers[6], quantum cryptography[7], and quantum memories[8].

It is important to note that the interaction of structured light with matter drastically changes depending on the degree of size mismatch between them[9]. While structured light can be easily generated and manipulated on the macro scale, modern applications of OAM beams for subwavelength imaging, high-capacity communication systems, and micromanipulation motivate the development of ultra-compact, microscale sources of light with SAM and/or OAM. When it comes to the interactions of light and matter at the atomic level, for example, the probing of atomic forbidden transitions (or dark states)[10] and building multi-dimensional quantum encryption systems[11, 12], the scale of the beam has to be reduced down to the nanoscale to be comparable to the size of quantum dots or even atoms. Indeed, many light-matter interaction processes, such as dipole-forbidden spectroscopic transitions are highly improbable. A majority of previously (mostly theoretically) proposed methods[13-15] relied on designing the structure around the emitter to modify the selection rules. For example, Heeres and Zwiller theoretically proposed a plasmonic nano-antenna-based method for enhancing the light-matter interaction strength by creating a phase singularity with a sub-diffraction limit scale using antennas[16].

On the other hand, theoretical studies in solid-state physics predicted that OAM beams can induce electric currents and secondary magnetic fields in the bulk structures and quantum rings[17], new electronic transitions (forbidden for plane waves) in quantum dots[9], deflection of electron wave packets in semiconductors[18], new magnetic phenomena close to the phase singularity[19], and new transitions at normal incidence (forbidden for plane waves) in nanostructures[20]. Notably, the transition rate depends on the vortex beam size. As the beam size decreases the transition rate increases[21].

## Results

Here we experimentally demonstrate a new approach for de-magnifying the OAM beams to a deep subwavelength scale using a flat hyperbolic metamaterial (HMM) with extreme anisotropy of dielectric permittivity characterized by a hyperbolic dispersion relation. The strong focusing capacity of the proposed HMM structure, referred to as hypergrating in this work, originates from the presence of the propagating modes with high in-plane wave numbers facilitated by the HMM[22]. An illustration of the hypergrating is shown in Fig.1 (a), demonstrating circularly polarized light converted to a vortex beam that, in turn, is focused in the HMM.

We report experimental evidence of the OAM-carrying beam focused to approximately one-third of the vacuum wavelength $\lambda$ (about one-third of the spot size which can be achieved using a high-numerical-aperture objective lens[23]). The focusing behavior observed in experiments is consistent with theoretical predictions. Detailed analysis of light distribution at the focal plane of the hypergrating reveals the formation of an optical skyrmion, with deep subwavelength spin structures ($\lambda/250$). The proposed framework, applied here to the beam carrying OAM= $-1$, can be extended to higher-order OAM beams (see Supplementary information).

The internal structure of the OAM beam always contains phase singularities, regions of space where the light intensity is zero and phase is not defined, making OAM beams fundamentally different from their Gaussian counterparts. Such spatial structure, however, results in difficulty in focusing structured beams. Indeed, even the tightly focused OAM-carrying beams with topological charge 1 are typically (1.1 ×) larger than the free space wavelength[23].

These challenges can be alleviated when an OAM beam is focused within the HMMs that support the propagation of plane waves with extremely broad-wavenumber-spectra[24]. The focusing can be initiated with a Fresnel lens that is incorporated into a planar HMM slab, the combination known as a hypergrating [22]. In this work, the hypergrating is based on an HMM slab comprising alternating $Ti_3O_5$ and Ag layers. The in-plane and out-of-plane permittivity of the HMM slab can be estimated using effective medium theory (EMT), $\varepsilon_\parallel = f\varepsilon_1 + (1-f)\varepsilon_2$, $\varepsilon_\perp = \left(\frac{f}{\varepsilon_1} + \frac{1-f}{\varepsilon_2}\right)^{-1}$, where $f$ is the filling factor of $Ti_3O_5$, $\varepsilon_1 = 5.1$ and $\varepsilon_2 = -11.8 + 0.4i$ are the permittivity of $Ti_3O_5$ and Ag, respectively, at the wavelength of 532 nm. Here we use a 10-bilayer HMM stack with f = 0.5, resulting in $\varepsilon_\parallel = -3.3 + 0.2i$ and $\varepsilon_\perp = 18.2 + 0.4i$, indicating that the HMM operates in a type-II hyperbolic regime[25].

The optical response of the planar HMM was analyzed experimentally and compared with theoretical predictions (see Supplementary information). While EMT provides a reasonable insight into the physical behavior of the composite, the material parameters in the composites with relatively thick (~ 30 nm) layers may substantially deviate from the EMT predictions[26]. Here, we utilized the formalism describing photonics of periodically stratified media[27] to calculate the boundaries of the Fresnel zones across the planar HMM (see Supplementary information). A focusing hypergrating was then fabricated by blocking the even Fresnel zones with Cr film, as shown in Fig. 1 (b).

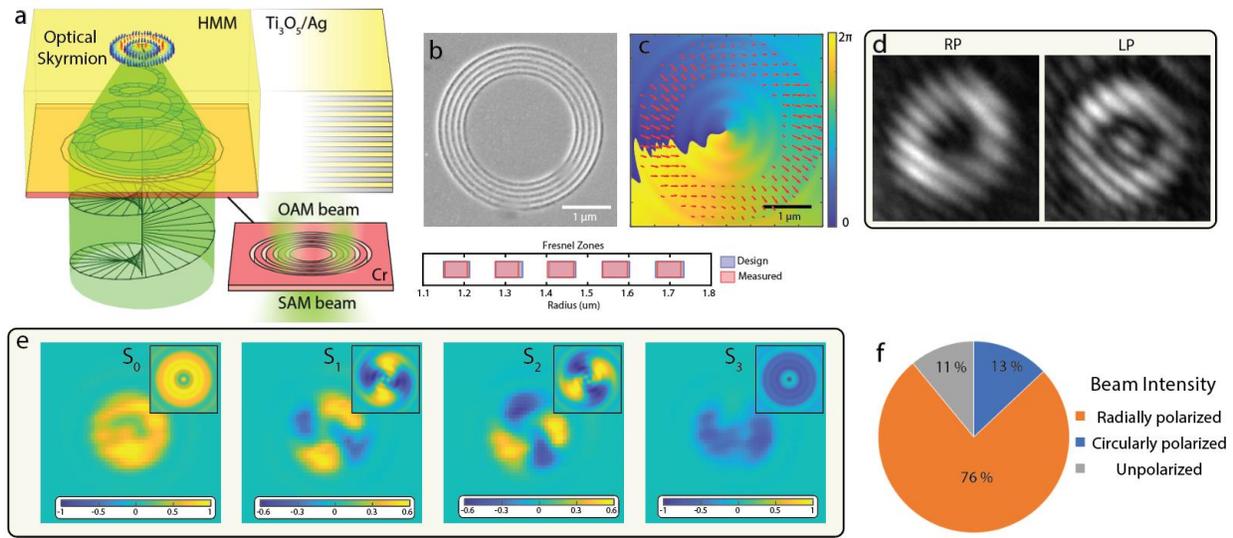

**Figure 1**. (a) Schematic diagram of a hypergrating structure and its working principle. (b) Scanning electron microscope (SEM) image of a Fresnel grating and the Fresnel grating rings' radiuses (designed and experimentally measured (c) Vector plot of the electric field (red arrows) in the transverse plane and the phase distribution of the radially polarized component ($E_r$) of the beam at 60 nm after the Fresnel grating. (d) Interference pattern of the left and right circularly polarized components of the output beam. (e) Experimentally measured Stokes parameters distributions in the far field (Inserted: Stokes parameters calculated from simulation). (f) Intensity distribution among components of the beam after the Fresnel grating.

Note that since the radial separation between the Fresnel zones is much smaller than a free-space wavelength, the Cr Fresnel grating transmits radially polarized light much more efficiently than azimuthally polarized light. Therefore, when a circularly polarized beam (or a beam with SAM) is incident on this grating, it is effectively converted into the radially polarized OAM beam at the exit of the grating. It is important to note that the output beam may still contain a fraction of circularly polarized light due to the incomplete absorption of the azimuthally polarized component of the input beam. To estimate the relative intensity fractions of the OAM beam and circularly polarized SAM beam in the output beam, we performed finite-element simulation with COMSOL for a circularly polarized beam passing through the Fresnel grating. The electric field vectors of the output beam are mainly oriented along the radial direction, indicating the output beam is predominantly radially polarized which carries a helical wavefront and OAM (Fig. 1 (c))[28]. Based on the simulation results, it has been determined that the OAM beam constitutes approximately 83% of the total intensity, while the remaining 17% corresponds to the circularly polarized beam.

The SAM to OAM conversion was confirmed by acquiring the phase information of the output beam using interference measurement. While direct interference measurement of a polarized beam is tricky, the radially polarized beam with topological charge $l = -1$ can be represented as a linear combination of left and right circularly polarized beams with charge $l = 0$ and $l = -2$ . $[\cos(\varphi), \sin(\varphi)]e^{-i\varphi} = 0.5[1, -i] + 0.5[1, i]e^{-2i\varphi}$ [29]. The total topological charge of the left and right circularly polarized components was measured to be 0 and -2 as is shown in Fig. 1 (d), confirming the charge -1 OAM beam existing in the output beam.

The polarization state of the output beam was measured experimentally to determine the OAM beam contribution as described in Ref. 30, 31 and in the Supplementary information. The results are depicted in Fig. 1 (e) in the form of Stokes parameters distributions. The portion of radially polarized OAM beam accounts for $76 \pm 1\%$ of total beam intensity (see Supplementary information for details). 11% of the beam lost its polarization state after passing through the Fresnel grating, an effect that could be attributed to scattering.

Once the generation of the OAM beam is confirmed, the Fresnel grating has been incorporated into a planar HMM, resulting in the hypergrating structure. The light propagation through the hypergrating has been analyzed numerically and experimentally.

Figure 2 (a) shows the predicted performance of the structure, illuminated by a circularly polarized beam with a 532 nm wavelength. The intensity distribution at 15 nm above the HMM top surface demonstrates the full width at half maximum of the focused beam to be about 110 nm or around $\lambda/5$. The same hypergrating supports tight focusing of vortex beams that carry higher OAM charge, surpassing the performance achieved with high numerical aperture (NA) objectives (see Supplementary information). The advanced focusing ability stems from the propagation of the high in-plane wave number modes supported by the HMM, which is not feasible in mediums such as air or immersion oil. Note that due to the total internal reflection at the top surface, the focused beams cannot be directly outcoupled to free space but instead become an evanescent beam above the HMM.

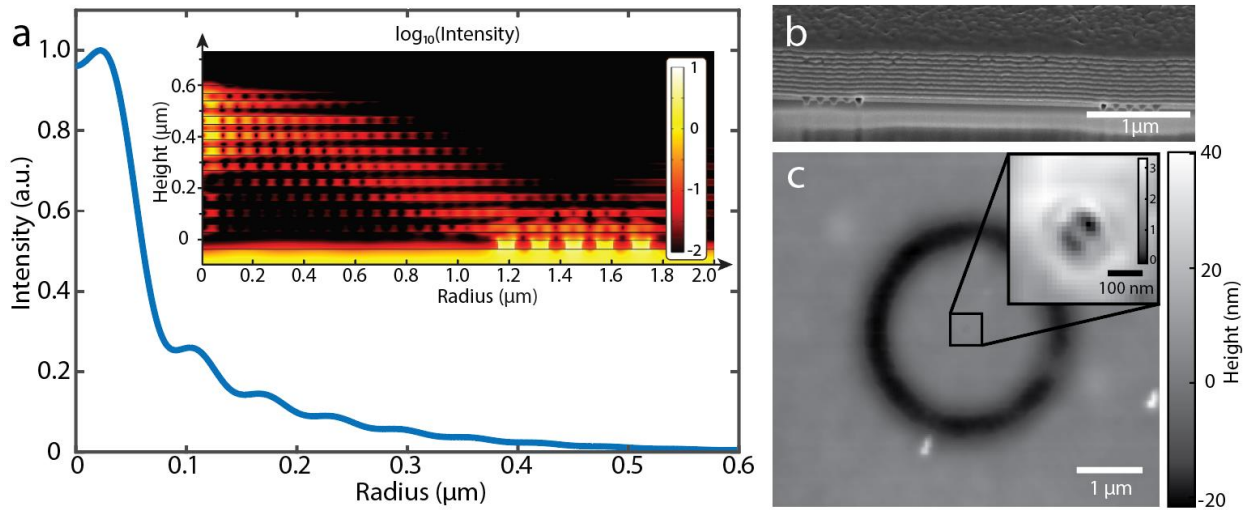

**Figure 2.** (a) Light intensity distribution at the focus of the hypergrating (10 nm above the HMM) from numerical simulation. The inserted figure shows the light intensity distribution in the HMM. Note that the intensity distribution has axial symmetry with respect to the principal axis. (b) Cross section of the hypergrating structure. (c) The surface topology of the Azo-polymer after the exposure measured by atomic force microscopy.

An SEM image of the cross-section of the hypergrating structure is shown in Fig. 2 (b) (see Supplementary information for details of the fabrication procedure). To experimentally characterize the near-field intensity distribution in the focal plane of the hypergrating, we employed an Azo-polymer specifically designed for 532 nm light as the light detection material. The Azo-polymer undergoes photoisomerization cycles between the Trans and Cis states under illumination, resulting in the formation of a nano-scale surface relief pattern at the illuminated area[32]. Figure 2 (c) shows the results of the surface topology of the Azo-polymer film upon the 5-min exposure of hypergrating structure to 216 W/cm$^2$ with 532 nm continuous wave laser illumination. Besides the approximately 20-nm-thick donut-shaped valley that already existed before illumination (see Supplementary information for details), a figure-8-shaped surface relief pattern can be seen at the focal point of the hypergrating that appeared after the hypergrating sample was exposed to 532-nm light. This pattern results from the illumination of the tightly focused OAM beam and was measured to be about 200 nm in diameter and 3 nm from hilltop to valley, which suggests that the size of the focused beam is around 200 nm, or $\lambda/2.7$. While the AFM measurements suggest that the focusing spot is larger than theoretical predictions $\lambda/5$, it needs to be noted that the surface relief pattern might not precisely reflect the light intensity distribution due to the photoisomerization of Azo-polymer. In addition, the optical response of the fabricated sample is affected by the small roughness and non-planarity of the layers.

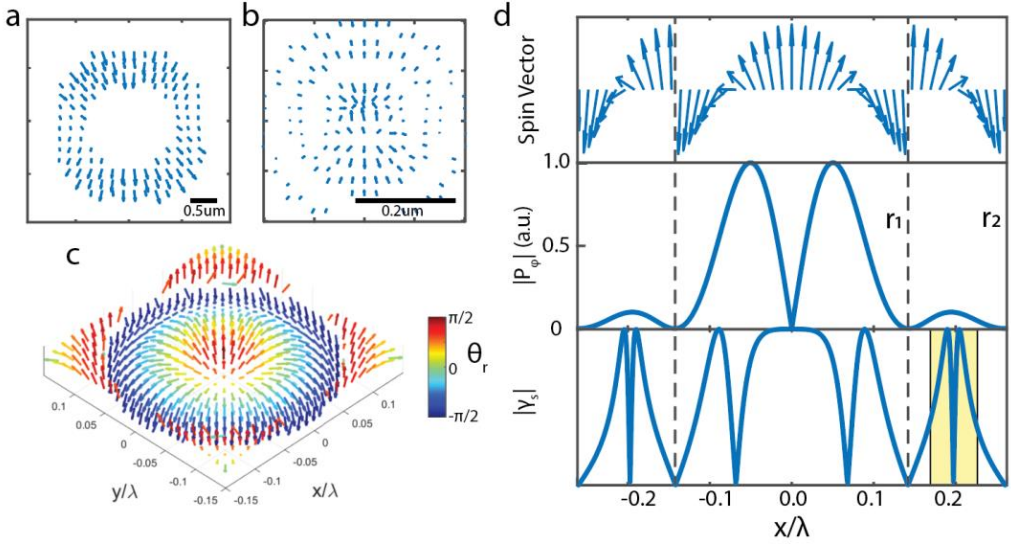

**Figure 3.** (a, b) Electric field distribution at 45 nm above the Cr layer and 15 nm above the HMM layer, respectively. (c) Unit spin vector distribution at 15 nm above the HMM, the color represents the $\theta_r = \tan^{-1}(S_z/S_r)$. (d) Spin vector (top), Poynting vector $P_\varphi$ (middle), and spin state $\gamma_s$ (bottom) variation across the center of the beam at 15nm above the HMM.

Next, we analyze the structure of the EM field in the vicinity of the focal point of the hypergrating. As seen in Fig. 3(a), the electric field (45 nm) above the hypergrating preserves its radial polarization. At the same time, the distribution at the focal plane (15 nm above the HMM surface, Fig.3b) exhibits a complex polarization structure, indicating strong spin-orbit interaction. Detailed analysis of the distribution reveals that such spin-orbit interaction results in a Neel-type skyrmion spin vector distribution forming in the evanescent field. The SAM density is defined as $S = Im(\varepsilon E^* \times E + \mu H^* \times H)/4\omega$, where $E$ and $H$ are the electric and magnetic fields, $\varepsilon$ and $\mu$ is the free space permittivity and permeability, and $\omega$ is the frequency. The spin vector is the unit vector of the SAM density ($\sigma = S/|S|$). To better illustrate the spin structure at the focal plane, the spin vector $\sigma$ of the evanescent wave at 15 nm above the top hypergrating surface is shown in Fig. 3 (c). The spin vector reverses from the "up" to the "down" state along the radial direction with a flip at the "domain wall", where the Poynting vector $P_\varphi = 0$ as is shown in Fig. 3 (d) [33]. The reversal of the spin vector within the domains occurs at deep sub-wavelength scales. The local spin structure defined as $\gamma_s = \frac{I_{RCP}-I_{LCP}}{I_{RCP}+I_{LCP}}$ [33] is shown in Fig. 3 (d). In particular, the spin-flip within the first and second domains occurs within of 8 nm ($\lambda/60$) and 2 nm ($\lambda/250$) respectively. Optical skyrmions hold the promise of unveiling novel light-matter interactions within quantum chiral materials, potentially opening a wide array of applications that span from high-resolution imaging and precision metrology to the realms of quantum technologies and advanced data storage methods[33].

## Conclusion

To conclude, we theoretically demonstrated a hypergrating system that focuses a vortex beam below the diffraction limit and supports optical skyrmion formation. Numerical simulation was employed to study the light propagation in the hypergrating and predicted that the radially polarized vortex beam can be focused down to $\lambda/5$ owing to the high in-plane wave number modes supported by the hypergrating, which is ~5 times tighter focusing than achievable with a 0.985-NA objective lens. The spin vector at the focal plane of the hypergrating behaves as a skyrmion with deep subwavelength spin structures ($\lambda/250$). Experimentally, the hypergrating structure was fabricated and the beam size was measured to be around $\lambda/3$ at the focal spot. The significant enhancement in focusing power holds the potential to facilitate the transfer of OAM to matter and enable the exploration of novel light-matter interaction phenomena.

**References**

1. Padgett, M.J. Orbital angular momentum 25 years on [Invited]. *Opt Express* **25**, 11265-11274 (2017).
2. Franke-Arnold, S., Allen, L. & Padgett, M. Advances in optical angular momentum. *Laser Photonics Rev* **2**, 299-313 (2008).
3. Bliokh, K.Y., Rodríguez-Fortuño, F.J., Nori, F. & Zayats, A.V. Spin-orbit interactions of light. *Nat Photonics* **9**, 796-808 (2015).
4. Schmiegelow, C.T. et al. Transfer of optical orbital angular momentum to a bound electron. *Nat Commun* **7** (2016).
5. Fickler, R. et al. Interface between path and orbital angular momentum entanglement for high-dimensional photonic quantum information. *Nat Commun* **5** (2014).
6. Curtis, J.E. & Grier, D.G. Modulated optical vortices. *Opt Lett* **28**, 872-874 (2003).
7. Mirhosseini, M. et al. High-dimensional quantum cryptography with twisted light. *New J Phys* **17** (2015).

8. Nicolas, A. et al. A quantum memory for orbital angular momentum photonic qubits. *Nat Photonics* **8**, 234-238 (2014).
9. Quinteiro, G.F. & Kuhn, T. Light-hole transitions in quantum dots: Realizing full control by highly focused optical-vortex beams. *Phys Rev B* **90** (2014).
10. Rivera, N., Kaminer, I., Zhen, B., Joannopoulos, J.D. & Soljacic, M. Shrinking light to allow forbidden transitions on the atomic scale. *Science* **353**, 263-269 (2016).
11. Fang, X.Y., Ren, H.R. & Gu, M. Orbital angular momentum holography for high-security encryption. *Nat Photonics* **14**, 102-+ (2020).
12. He, C., Shen, Y.J. & Forbes, A. Towards higher-dimensional structured light. *Light-Sci Appl* **11** (2022).
13. Lembessis, V.E. & Babiker, M. Enhanced Quadrupole Effects for Atoms in Optical Vortices. *Phys Rev Lett* **110** (2013).
14. Afanasev, A., Carlson, C.E. & Mukherjee, A. High-multipole excitations of hydrogen-like atoms by twisted photons near a phase singularity. *J Optics-Uk* **18** (2016).
15. Mahdavi, M., Sabegh, Z.A., Mohammadi, M., Hamedi, H.R. & Mahmoudi, M. Manipulation and exchange of light with orbital angular momentum in quantum-dot molecules. *Phys Rev A* **101** (2020).
16. Heeres, R.W. & Zwiller, V. Subwavelength Focusing of Light with Orbital Angular Momentum. *Nano Lett* **14**, 4598-4601 (2014).
17. Quinteiro, G.F. & Berakdar, J. Electric currents induced by twisted light in Quantum Rings. *Opt Express* **17**, 20465-20475 (2009).
18. Reinhardt, O. & Kaminer, I. Theory of Shaping Electron Wavepackets with Light. *Acs Photonics* **7**, 2859-2870 (2020).
19. Quinteiro, G.F., Reiter, D.E. & Kuhn, T. Formulation of the twisted-light-matter interaction at the phase singularity: Beams with strong magnetic fields. *Phys Rev A* **95** (2017).
20. Sedeh, H.B. et al. Manipulation of Scattering Spectra with Topology of Light and Matter. *Laser Photonics Rev* **17** (2023).
21. Schmiegelow, C.T. & Schmidt-Kaler, F. Light with orbital angular momentum interacting with trapped ions. *Eur Phys J D* **66** (2012).
22. Thongrattanasiri, S. & Podolskiy, V.A. Hypergratings: nanophotonics in planar anisotropic metamaterials. *Opt Lett* **34**, 890-892 (2009).
23. Rao, L.Z., Pu, J.X., Chen, Z.Y. & Yei, P. Focus shaping of cylindrically polarized vortex beams by a high numerical-aperture lens. *Opt Laser Technol* **41**, 241-246 (2009).
24. Cortes, C.L., Newman, W., Molesky, S. & Jacob, Z. Quantum nanophotonics using hyperbolic metamaterials. *J Optics-Uk* **14** (2012).
25. Poddubny, A., Iorsh, I., Belov, P. & Kivshar, Y. Hyperbolic metamaterials. *Nat Photonics* **7**, 948-957 (2013).
26. Elser, J., Podolskiy, V.A., Salakhutdinov, I. & Avrutsky, I. Nonlocal effects in effective-medium response of nanolayered metamaterials. *Appl Phys Lett* **90** (2007).
27. Avrutsky, I. Guided modes in a uniaxial multilayer. *J Opt Soc Am A* **20**, 548-556 (2003).
28. Zhan, Q.W. Cylindrical vector beams: from mathematical concepts to applications. *Adv Opt Photonics* **1**, 1-57 (2009).
29. Liu, Y.F. et al. Optical Vector Vortex Generation by Spherulites with Cylindrical Anisotropy. *Nano Lett* **22**, 2444-2449 (2022).
30. Singh, K., Tabebordbar, N., Forbes, A. & Dudley, A. Digital Stokes polarimetry and its application to structured light: tutorial. *J Opt Soc Am A* **37**, C33-C44 (2020).
31. Chen, S.Z. et al. Generation of arbitrary cylindrical vector beams on the higher order Poincare sphere. *Opt Lett* **39**, 5274-5276 (2014).

32. Masuda, K. et al. Azo-polymer film twisted to form a helical surface relief by illumination with a circularly polarized Gaussian beam. *Opt Express* **25**, 12499-12507 (2017).
33. Du, L.P., Yang, A.P., Zayats, A.V. & Yuan, X.C. Deep-subwavelength features of photonic skyrmions in a confined electromagnetic field with orbital angular momentum. *Nat Phys* **15**, 650-+ (2019).

**Acknowledgments**

**Funding:** N.M.L. would like to acknowledge the support of this research by the Army Research Office Award no. W911NF2310057. V.A.P. would like to acknowledge the support of this research by the National Science Foundation, Division of Materials Research Award no. DMR 2004298. **Author contributions:** W.L., J.L., E.S., V.A.P., and N.M.L. designed, conducted the simulations and experiments. W.L., A.C., R.Y.B, J.D.M, J.F., and N.M.L. fabricated the Hypergrating samples. T.O. provided the Azo-polymers. All authors analyzed the results and edited the paper. **Competing interests:** Authors declare that they have no competing interests.

# Nano-Focusing of Vortex Beams with Hyperbolic Metamaterials

# Supplementary Information

Wenhao Li[1], Jacob LaMountain[2], Evan Simmons[2], Anthony Clabeau[3], Robel Y. Bekele[3], Jason D. Myers[3], Takashige Omatsu[4], Jesse Frantz[3], Viktor A. Podolskiy[2], Natalia M. Litchinitser[1*]

[1]Department of Electrical and Computer Engineering, Duke University; Durham, NC, 27708, USA

[2]Department of Physics and Applied Physics, University of Massachusetts Lowell; Lowell, MA, 01854, USA

[3]US Naval Research Lab; Washington, DC, 20375, USA

[4]Molecular Chirality Research Center, Chiba University; Chiba, 263-8522, Japan

* Corresponding author. Email: natalia.litchinitser@duke.edu

## 1. Optical response of planar HMM

The optical response of the planar hybrid metamaterial (HMM) was studied using spectroscopic ellipsometry measurements. The change in polarization state upon reflection, represented as Ψ and Δ, was analyzed by measuring at incident angles of 65°, 70°, and 75°. To determine the thicknesses of the Ag and $Ti_3O_5$ layers, the measured Ψ and Δ curves were fitted to the model predicted using transfer matrix method given the refractive indexes of the individual Ag and $Ti_3O_5$ layers.

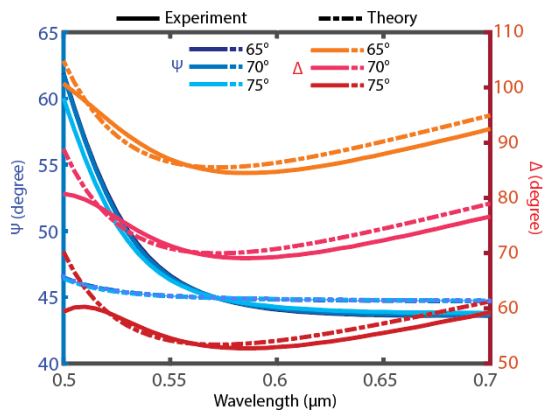

**Figure S1.** Ellipsometry measurement of the 10-bilayer HMM in reflection mode.

Figure S1 depicts the experimentally measured Ψ and Δ curve as a function of wavelength fitted to a 10-bilayer model with the thickness of Ag and $Ti_3O_5$ as fitting parameters. The fitting model for Ψ and Δ assumes identical thicknesses for all Ag layers and for all $Ti_3O_5$ layers. The determined thickness values from the fitting process are as follows: Ag layers have a thickness of 28.8 nm, and $Ti_3O_5$ layers have a thickness of 32.0 nm.

## 2. Fresnel Grating Design

To realize sub-diffraction focusing using the high in-plane wavenumber modes, we engineered the light propagation to realize positive interference at the output plane (corresponding to the focal point) of the hypergrating. The input plane of the hypergrating was divided into concentric rings based on the phase acquired by the light during its propagation from the input plane to the focal point. However, the light propagation in the HMM is not well described by the effective medium theory when the layers in the HMM are relatively thick (in our case, they are 30 nm). Therefore, we describe modes supported by HMM using expressions developed for periodically stratified media [26]. The dispersion of modes (the dependence of the components of the wavevector $k_x, k_z$ on the angular frequency $\omega$) is given by:

$$f(k_x, k_z, \omega) = \cos(2k_z a) - \cos(k_1 a)\cos(k_2 a) + \gamma \sin(k_1 a)\sin(k_2 a) = 0,$$

where $a$ is the layer thickness, $\gamma$ is a polarization-specific parameter

$$\gamma_{TM} = \frac{1}{2}\left(\frac{\varepsilon_2 k_1}{\varepsilon_1 k_2} + \frac{\varepsilon_1 k_2}{\varepsilon_2 k_1}\right), \gamma_{TE} = \frac{1}{2}\left(\frac{\varepsilon_2}{\varepsilon_1} + \frac{\varepsilon_1}{\varepsilon_2}\right), \text{ and } k_{1,2}^2 = \varepsilon_{1,2}\frac{\omega^2}{c^2} - k_x^2$$

The direction of the pulse propagating in the layered media can be calculated by analyzing the Poynting vector. In the loss-less limit,

$$S_x \propto \frac{\partial \omega}{\partial k_x} = \frac{\partial f}{\partial k_x}\left(\frac{\partial f}{\partial \omega}\right)^{-1}, S_z \propto \frac{\partial \omega}{\partial k_z} = \frac{\partial f}{\partial k_z}\left(\frac{\partial f}{\partial \omega}\right)^{-1}$$

The design of the grating employs the above relationship between the direction of the wavevector (phase velocity) and that of the beam propagation (group velocity) along with the requirement for the beams from the neighboring Fresnel zones to arrive at the focal point with a phase shift of $\pi$. Assuming the normally-incident plane wave field at the grating and the target focal length $f_0$, the edges of the Fresnel zones are given by:

$$\left(k_{x_i} x_i + k_{z_i} f_0\right) - \left(k_{x_0} x_0 + k_{z_0} f_0\right) = m\pi,$$

where $x_j$ is the position of $j$-th Fresnel zone edge and $\{k_{x_j}, k_{z_j}\}$ represent the components of the wavevector of the beam propagating from the edge of the zone to the focal point. Note that for type II HMMs, used in this work, $x_0$ needs to be larger than the cutoff range, which is calculated to be 1.07 μm. The optimization of the $x_0$ position is performed with simulations. The best focusing is achieved at $x_0 = $ 1.10 μm in the lossless approximation and $x_0 = 0.88$ μm with losses. Experimentally the best focusing was achieved for $x_0 = 1.15$ μm. The difference in the experiment and the theory might be due to the roughness-induced scattering or inaccuracies in the film thickness and refractive indexes.

### 3. SAM to OAM conversion

Circularly polarized beams can be decomposed into a combination of radially polarized and azimuthally polarized components with spiral phase wavefront. Here, the period of the gratings is much smaller than the incident beam wavelength, and the transmitted beam is predominantly radially polarized if the azimuthal component of the field is strongly absorbed.

$$\vec{E_{in}} = E_0(\vec{e_x} - i\vec{e_y}) = E_0 e^{-i\phi}(\vec{e_r} - i\vec{e_\phi}),$$

$$\vec{E_{out}} = E_r e^{-i\phi}\vec{e_r} - iE_\phi e^{-i\phi}\vec{e_\phi},$$

where $|E_\phi|$ is much smaller than $|E_r|$. The output beam becomes a predominantly radially polarized vortex beam due to the conversion of the spin angular momentum to orbital angular momentum. The output beam can be represented as a combination of radially polarized component and circularly polarized component for easy interpolation and for quantitative analysis (for details see part 5 of the supplementary information):

$$\vec{E_{out}} = (E_r - E_\phi)e^{-i\phi}\vec{e_r} + (E_\phi e^{-i\phi}\vec{e_r} - iE_\phi e^{-i\phi}\vec{e_\phi}) = E'_r e^{-i\phi}\vec{e_r} + E_\phi e^{-i\phi}(\vec{e_r} - i\vec{e_\phi})$$
$$= E'_r e^{-i\phi}\vec{e_r} + E_\phi e^{-i\phi}(\vec{e_x} - i\vec{e_y})$$

### 4. Numerical simulation

Numerical simulations were performed using COMSOL Multiphysics. The simulation setup has axial symmetry, and the input beam is a 532 nm wavelength circularly polarized plane wave. The hypergrating structure is designed on a glass substrate, consisting of a Cr Fresnel grating layer followed by 20 alternating layers of 30 nm thick Ag and Ti3O5. All the refractive indexes used in the simulation were experimentally measured with ellipsometry. The Fresnel zone positions were calculated using the dispersion relation of the HMM.

### 5. Stokes parameters analysis

The Stokes parameters measurement was based on four intensity measurements with the setup shown in Figure S2 known as reduced Stokes polarimetry[30]. The output beam was collected with an objective (Nikon, 100x, 0.8NA), passed through the quarter-wave plate (QWP) and a linear polarizer (LP), and then recorded with a camera (Princeton Instrument, Pixis). The circular intensity profiles ($I_R$ and $I_L$) were acquired by setting the QWP at 90 degrees and the LP at 45 and 135 degrees respectively. By removing the QWP and adjusting the polarizer to angular orientations of 0 and 45 degrees, the two linear intensities ($I_D$ and $I_H$) were acquired. The Stokes parameters were obtained using the following equations: $S_0 = I_R + I_L$, $S_1 = 2I_H - S_0$, $S_2 = 2I_D - S_0$, $S_3 = I_R - I_L$.

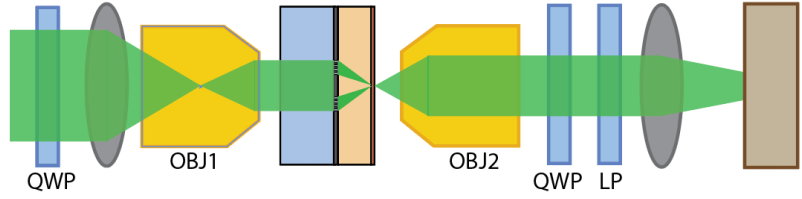

**Figure S2.** Experimental setup for measuring the Stokes parameters.

The output beam of the Fresnel grating can be represented as a linear combination of a radially polarized vortex beam and a circularly polarized beam,

$$\overrightarrow{E_{out}} = E_{r0}e^{-i\phi}\overrightarrow{e_r} + \frac{1}{\sqrt{2}}E_{s0}(\overrightarrow{e_x} - i\overrightarrow{e_y}),$$

where $E_{r0}$ and $E_{s0}$ are the electric field distribution of the vortex and circularly polarized beam components. From the generalized form of Stokes parameters[30], the output beam of the Fresnel grating has the Stokes parameters with the following form:

$$S_0 = E_{r0}^2 + E_{s0}^2 + \sqrt{2}E_{r0}E_{s0}$$

$$S_1 = (E_{r0}^2 + \sqrt{2}E_{r0}E_{s0})\cos(2\phi)$$

$$S_2 = (E_{r0}^2 + \sqrt{2}E_{r0}E_{s0})\sin(2\phi)$$

$$S_3 = E_{s0}^2 + \sqrt{2}E_{r0}E_{s0}$$

$S_1$ and $S_2$ distributions have four quarters and are rotated by 45 degrees. The Stokes parameters distributions $S_0$, $S_1$, $S_2$, and $S_3$ from experimental measurement and simulation are shown in the main paper Fig. 1 (e). From $S_1$, $S_2$, and $S_3$, the magnitude of the $E_{s0}$ and $E_{r0}$ distributions can be calculated.

Figure S3 (a) and (b) show the $E_{s0}^2$ and $E_{r0}^2$ distribution of a circularly polarized beam passed through the Fresnel gratings. The total intensity of the intensity of the vortex beam and circularly polarized are determined using the following equations,

$$I_v = \iint E_{r0}^2 dxdy, \; I_s = \iint E_{s0}^2 dxdy$$

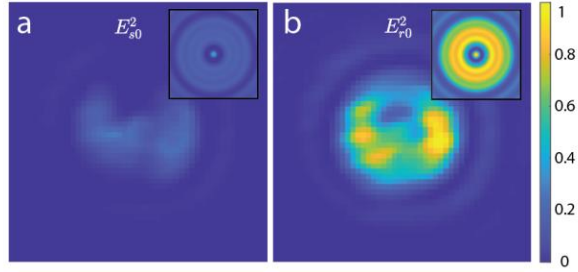

**Figure S3.** $E_{s0}^2$ and $E_{r0}^2$ distributions calculated from the Stokes parameters.

In the simulation results, the circularly polarized beam accounts for 17% of the total power, while the radially polarized vortex beam constitutes the remaining 83%. In the experiment, the degree of polarization, defined as $p = \frac{I_{pol}}{I_{tot}} = \frac{\sqrt{S_1^2 + S_2^2 + S_3^2}}{S_0}$ is measured to be 0.89. The contributions to the total intensity from the unpolarized beam, circularly polarized beam, and radially polarized vortex beam are 11%, 13%, and 76%, respectively. The loss of polarization might be due to scattering.

## 6. Focusing high-order OAM beam with Hypergrating

The focusing behavior of a high OAM vortex beam with the hypergrating device was studied using COMSOL simulations. The input beam is a radially polarized beam defined as $E_{in} = E_0(r)\, e^{il\varphi} \vec{e_r}$, where charges $l$ is the total topological charge, and $E_0$ is a constant. The intensity distribution at 15 nm above the HMM is shown in Figure S4.

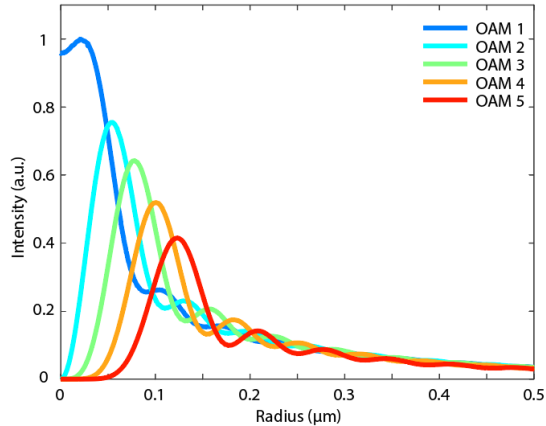

**Figure S4.** intensity distribution at 15 nm above the HMM for beams with topological charge 1~5.

With higher OAM the focused spot size is also getting larger. Here we define the spot size using the diameter of the donut shape, the diameter is set to be the position where intensity drops to half of the maximum. The focused spot sizes for beams carrying OAM charge 1~5 are 120 nm, 170 nm, 220 nm, 270 nm, and 310 nm respectively.

## 7. Hypergrating fabrication

The hypergrating fabrication process is shown in Fig. S5. The Fresnel gratings were fabricated by milling the designed concentric rings on a 50 nm thick Cr layer with a focused ion beam. The Fresnel gratings were filled with PMMA for planarization. Around 250 nm PMMA layer was spin-coated on top of the Fresnel gratings and the bulk area above the Cr surface was removed with reactive ion etching.

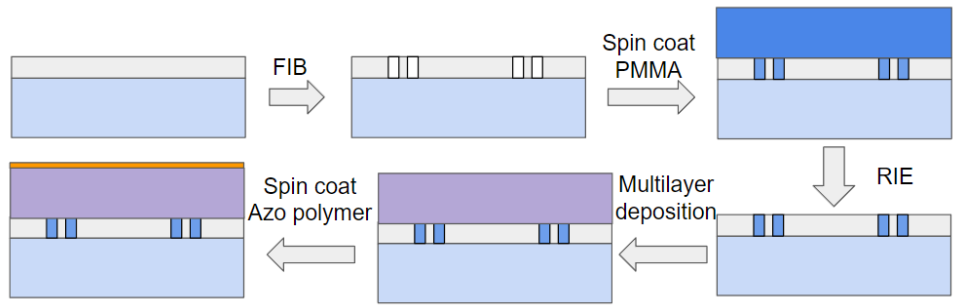

**Figure S5.** Hypergrating fabrication procedures.

On top of the Fresnel grating, 20 alternating layers of 30 nm thick Ag and $Ti_3O_5$ were deposited using E-beam evaporation. The thicknesses of the layers were calibrated with ellipsometry measurements.

## 8. Azo-polymer surface topology characteristics

Since the beam was confined within the hypergrating and only the evanescent wave was above the HMM, a near-field method was required to probe the focused beam. We used the surface relief pattern of the Azo-polymer to map out the area under illumination.

A 40 nm thick layer of Azo-polymer was spin-coated on top of the hypergrating to probe the intensity distribution of the evanescent field. A circularly polarized 532 nm laser was focused on the back focal plane of the incident objective to create a plane wavefront at the input of the hypergrating sample while the sample can be imaged with the same objective as is shown in Figure S2.

The hypergrating was placed under 216 W/cm$^2$ illumination for 5 minutes and the surface topology of the Azo-polymer was measured with atomic force microscopy. The surface topology of the Azo-polymer before and after illumination is shown in Fig. S6 (a) and (b).

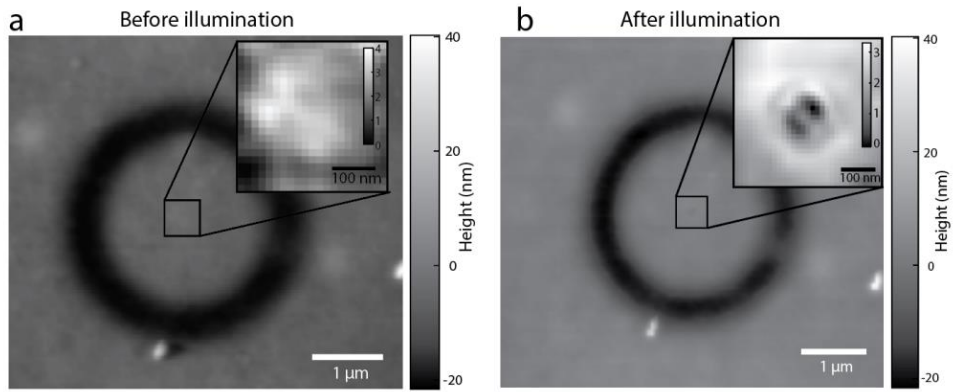

**Figure S6.** Surface topology of the Azo-polymer before and after exposure.

Note that a donut-shaped valley was observed before and after illumination at the position of the Fresnel gratings and is about 20 nm deeper than the surroundings. This valley resulted from the non-planarity of the hypergrating device which can also be observed in the hypergrating cross-section plot (main paper, Fig 2(b)). After exposure, an additional figure-8-shaped surface relief pattern becomes apparent at the focal point of the hypergrating. This pattern arises due to the illumination of the focused beam from the hypergrating.

## References

1. Elser J, Podolskiy VA, Salakhutdinov I, Avrutsky I. Nonlocal effects in effective-medium response of nanolayered metamaterials. Appl Phys Lett. 2007;90(19).

2. Singh K, Tabebordbar N, Forbes A, Dudley A. Digital Stokes polarimetry and its application to structured light: tutorial. J Opt Soc Am A. 2020;37(11):C33-C44.